# Pilot Low-Cost Concentrating Solar Power Systems Deployment in Sub-Saharan Africa: A Case Study of Implementation Challenges

Emmanuel Wendsongre Ramde [1,2], Eric Tutu Tchao [3]*, Yesuenyeagbe Atsu Kwabla Fiagbe [2], Jerry John Kponyo [4] and Asakipaam Simon Atuah [4]

[1] The Brew-Hammond Energy Centre, Kwame Nkrumah University of Science and Technology, Kumasi, 00233, Ghana; ewramde.soe@knust.edu.gh
[2] Department of Mechanical Engineering, Kwame Nkrumah University of Science and Technology, Kumasi, 00233, Ghana; yakfiagbe.soe@knust.edu.gh
[3] Department of Computer Engineering, Kwame Nkrumah University of Science and Technology, Kumasi, 00233, Ghana
[4] Department of Telecom. Engineering, Kwame Nkrumah University of Science and Technology, Kumasi, 00233, Ghana; jjkponyo.soe@knust.edu.gh (J.J.K.); simonasakipaam@gmail.com (A.S.A.)
* Correspondence: ettchao.coe@knust.edu.gh

**Abstract:** Electricity is one of the most crucial resources that drives any given nation's growth and development. The latest Sustainable Development Goals report indicates Africa still has a high deficit in electricity generation. Concentrating solar power seems to be a potential option to fill the deficit. That is because most of the components of concentrating solar power plants are readily available on the African market at affordable prices, and there are qualified local persons to build the plants. Pilot micro-concentrating solar power plants have been implemented in Sub-Saharan Africa and have shown promising results that could be expanded and leveraged for large-scale electricity generation. An assessment of a pilot concentrating solar power plant in the sub-region noticed one noteworthy obstacle that is the failure of the tracking system to reduce the operating energy cost of running the tracking control system and improve the multifaceted heliostat focusing behavior. This paper highlights the energy situation and the current development in concentrating solar power technology research in Africa. The paper also presents a comprehensive review of the state-of-the-art solar tracking systems for central receiver systems to illustrate the current direction of research regarding the design of low-cost tracking systems in terms of computational complexity, energy consumption, and heliostat alignment accuracy.

## 1. Introduction

Electrical energy is one of the most crucial resources that drives every given nation's growth and development [1,2]. However, as the commonest and most sustainable forms of electrical energy today, fossil fuels are central in their source of production. Moreover, fossil fuels are limited in nature, and their reserve quantities are declining as a consequence of extensive use. Their use also has a detrimental impact on climate change [3].

Hence, countries around the world need to pursue sustainable sources of electrical energy production to meet their energy needs. This is particularly important for African developing countries embarking on programs designed to achieve economic growth and development. Research on the progress made in meeting Sustainable Development Goal 7, the ambitious goal of ensuring equal access to energy by 2030, shows that the number of people without access to power dropped from 1.2 billion in 2010 to 840 million in 2017 globally. The study on another scale estimates that by 2030, about 650 million people will still be without access to electricity and 9 out of 10 will live in sub-Saharan Africa (SSA). The drastic lack of access to energy suppresses economic growth and sustainable development [4]. Sadly, recent SSA policies and developments do not recognize the potentially tremendous social and economic advantages of access to electricity [5], nor do they consider the significant potential of solar off-grid and mini-grid technologies to provide clean energy. However, the solutions to the vast energy gap in SSA exist in the form of decentralized solar for electricity generation, among other renewable energy sources. Hence, there exists the need to accelerate investments and innovations in renewables aiming at reducing the disparity in access to electricity. Such developments could help satisfy the growing energy needs of the sub-region and the entire continent. Investments and developments in the renewable energy sector will also help to reduce the economic and social disadvantages that the resulting energy gap might bring. Renewable energy can also provide a vital solution that can help build the nearly non-existent and, in some situations, inadequate electricity infrastructure in many rural areas and even match the growth of the continent with the Sustainable Development Goal 7.

Due to its abundance and sustainability, solar energy is an essential resource of renewable energy in Africa [6]. The concentrating solar power (CSP) system is an important candidate for solar energy utilization in Africa. This is because CSP systems collect the direct solar radiation beam component making them best suited to areas with a high percentage of clear-sky days. Furthermore, the components needed for the construction of a CSP plant are readily available on the African market at low cost, and there is a qualified local human resource to build the plants [7]. CSP systems have been implemented in Africa, and the results demonstrate their great potential to expand the supply of electricity to rural communities [7]. The SDG 7 Policy Briefs, 2019 also revealed an increasing interest in investing in renewable energy in Africa [8].

However, initial pilot project developments on CSP systems have seen high cost of capital and low efficiency in operations. These factors have hindered their application on a broad scale [9]. Therefore, to make CSP technology economically viable in Africa, it is important to find ways to reduce the cost of its implementation and improve the efficiency of its operation. Major ways to cut costs are possible, especially for central receiver system (CRS) technologies where the area of heliostats accounts for 40%–50% of total capital expenditure [10] and the monitoring system is highly effective and computationally efficient [11]. The current developments, especially in sub-Saharan Africa, concentrate on cost reductions in assembly and manufacturing processes (such as the use of more local labor and materials to minimize plant construction costs), lighter mirrors, new and improved mirror materials, new heat transfer fluids such as molten salt and direct steam, and cost-effective tracking system design.

This paper seeks to evaluate a low-cost CSP system that was developed and implemented in Ghana to serve the electricity needs of a rural area in northern Ghana. The paper also analyzes and provides insight into similar CSP deployment in the sub-region and provides perspective on how to build and create these systems for optimum performance operation.

## 2. Background to CSP Technologies

Concentrating solar power (CSP) seems a potential option for the energy needs of African countries. CSP plants capture energy from solar radiation, converting it into heat to produce electricity using steam turbines, gas turbines, or Stirling engines [12]. CSP technology consists of both a solar portion (to focus and transform solar energy into usable thermal energy) and a conventional power block (to turn heat into electricity). The primary source of energy for CSP is direct normal irradiance (DNI), perpendicular to a surface that tracks the Sun continuously [13]. CSP plants have the maximum potential in the Earth's "sunbelt" [14]. There are various CSP technologies, but they have the same basic principle of electricity generation. The structure and focus of these systems are distinct, giving rise to different temperature ranges. The CSP systems are divided into line focusing and point focusing. The line focusing systems concentrate the solar radiation onto a linear absorber tube or series of tubes using the reflectors. The reflectors are then tracked about a single axis to keep the Sun's image focused onto the line absorber tube. Examples of line focusing systems include:

- Parabolic trough (PT): The parabolic trough concentrator system consists of several curved parallel mirrors that concentrate the Sun's rays on a receiver tube containing the heat transfer fluid [15]. The PT device has six main components that include surface reflecting, absorber, support structure, electricity generation unit, thermal storage unit, and tracking system. The tracking system tracks the Sun over the day along the central axis as the Sun moves east to west [16]. The tracking system can be designed for tracking the Sun by using photosensors on the PT concentrators or astronomical algorithms installed in the tracking system.

- Linear Fresnel reflector (LFR): The LFR utilizes flat-shaped Fresnel reflectors to concentrate the Sun's light on the receiver (linear absorbers) away from the system of reflection [17]. The LFR system also uses a single-axis tracking system to position the reflectors from east to west along a north–south axis to track the light.

- The point focusing systems: These systems concentrate the solar radiation at a central point called the receiver/tower. Examples include the solar dish (SD) system that uses parabolic dish-shaped solar concentrators that focus the sunlight onto a single central receiver [1].

The central receiver system (CRS) consists of a large number of heliostats, each of which adjusts its orientation periodically in two axes to cancel the Earth's movement relative to the Sun. This periodic heliostat adjustment continually directs the incident beams onto the fixed receiver above the tower [18]. This enables high-efficiency conversion of energy from solar to thermal at the receiver point. The heat transfer medium such as steam or molten salt carries away the thermal energy to power a generator for electricity generation or to be stored for later use [19]. Because of the receiver's fixed position, the heliostats typically do not point directly to the Sun resulting in a substantial reduction in the amount of solar radiation collected per unit area of the mirror relative to the Photo – Voltaic (PV) system. Moreover, the mirrors are positioned to avoid serious reflection of nearby mirrors as the Sun moves or blocking of some of the reflected light on its way to the receiver. This makes it necessary to forgo the collection of some of the incoming rays [20]. Figure 1 provides graphical descriptions of the various CSP technologies.

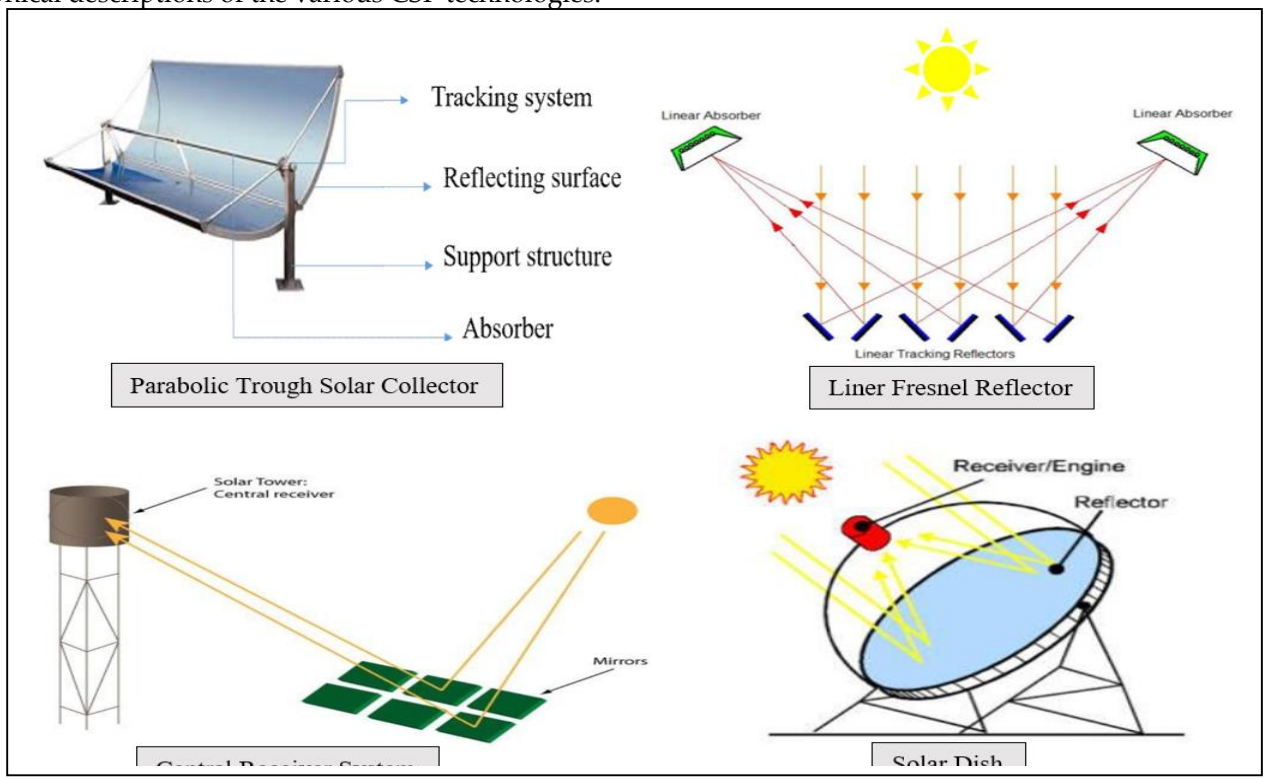

**Figure 1.** Concentrating solar power (CSP) technologies [2].

Unlike conventional power generation systems, the initial investment is about 80% of the cost of generating electricity under CSP [21]. CRS and Fresnel systems technologies are constructed from the same basic components as the PT technology. Fresnel systems have lower mirror and steel structure costs but this cost–benefit ratio is balanced by reduced plant performance due to lower optical efficiencies and lower working temperature. Even though, CRS requires an extra investment in the tower, it has lower piping and heliostats costs. It also has the benefit of producing higher temperatures and engaging local labor in the plant implementation. The remaining 20% represents the plant's operational and maintenance costs, a large portion of which the tracking system takes up. Within the CSP technology suite, the central receiver system offers lower costs, higher performance, greater participation of local labor, and dispatchable electricity generation, making it a much more preferred technology for sub-Saharan African countries.

## 3. State of the art of CSP Research and Development in Africa

Renewable energy sources (RESs) have gained widespread prominence in many African countries in alternative energy discussions. Electrification with RESs could address the continent's severe deficit in electricity supply, reduce energy security concerns, and alleviate the challenges posed by climate change [22]. CSP is one of several new renewable electricity generating technologies. However, CSP has high initial capital costs and complex deployment of technology and these serve as a major barrier to the implementation of CSP systems in many developing countries. Balghouthi et al [23] investigated solar resource potentials and the suitable factors for CSP implementation in Tunisia. The authors used data on solar radiation and weather parameter values from southern Tunisia to model a 50 MW solar power plant parabolic trough and compared the plant's energy and economic efficiency with a reference CSP plant in Spain (Andasol [24]). The results showed that Africa has very large solar resources appropriate for the deployment of CSP systems. The authors have estimated that the total annual generation of solar energy from the model plant exceeds the reference plant by 1793 MWhe. However, the projected total investment cost was higher than that of the reference plant. The authors concluded that the bulk of the plant components should be produced locally for CSP projects to be economically competitive in Africa. Aly et al [25] and Meligy et al. [17] made a similar observation when they modeled PT and CSP systems using the System Advisor Framework to investigate the techno-economic feasibility of CSP technology in Tanzania and Libya. The authors [25] achieved a levelized electricity cost (LCOE) of 11.6–12.5 c/kWh for the CSP model and 13.0–14.4 c/kWh for the PT under a debt interest rate of 7%.

In Nigeria, Olumide et al. [6] used data from a desktop survey to assess the most appropriate technology for solar thermal power plants and to evaluate CSP technologies under operating, environmental, and social conditions. The authors noted that because of its low cost and lesser land demand, PT is more appropriate, but it has no established commercial application. In comparison, it has been found that CSP is technically simpler and possesses better thermodynamic properties but has higher initial costs. Similar research work [26,27] was also carried out with the same aim of investigating the feasibility of CSP technology deployment in African countries.

Mihoub et al [28] investigated the optimal configuration and optimum construction of CSP plants in Algeria with minimal LCOE and maximum annual power generation. The authors introduced and contrasted different models and scenarios for PT and CRS using the LCOE. The authors concluded that CRS is the most attractive and desirable plant. Bouhal et al. [29] used parabolic trough collectors to evaluate the thermal performance of CSP systems in Morocco. The authors ran annual simulations in six climatic regions and found the outcome is strongly influenced by temperature and location. Elshafey et al. [30] presented a comprehensive review, several key issues, and a simple economic summary of the CSP system to address the issues of an optimized combined solar cycle in Egypt. The authors concluded that the techno-economic viability of CSP plants in Egypt is unquestioned. South Africa is considered to be one of the countries where CSP development is well-exploited. Craig et al. [31] used a system dynamics methodology to examine the unique, critical, and complex factors affecting CSP implementation in South Africa and observed that better research support is needed to open up new approaches and avenues to implement CSP technologies in South Africa. As part of a project consortium, Kwame Nkrumah University of Science and Technology Energy Center, the Brew-Hammond Energy Center (TBHEC), designed and built a cost-effective, compact, and dry CRS plant for an off-grid community in northern Ghana under CSP4Africa. This pilot CSP implementation was proof of concept and showed the feasibility of CSP in the sub-region. The other objective of this CSP4Africa project was to achieve a simple, robust, locally tailored, and sustainable low-cost CSP design for electricity generation for rural communities in the Sahelian region. The project has successfully demonstrated high involvement of local content and the use of low-cost local materials in implementing CSP plants. It also strengthened the local economy and capacity building and potentially lowered the local power cost for the off-grid community. The pilot plant also provides a platform for further research and skill development in CSP technology. Figure 2 shows the plant setup.

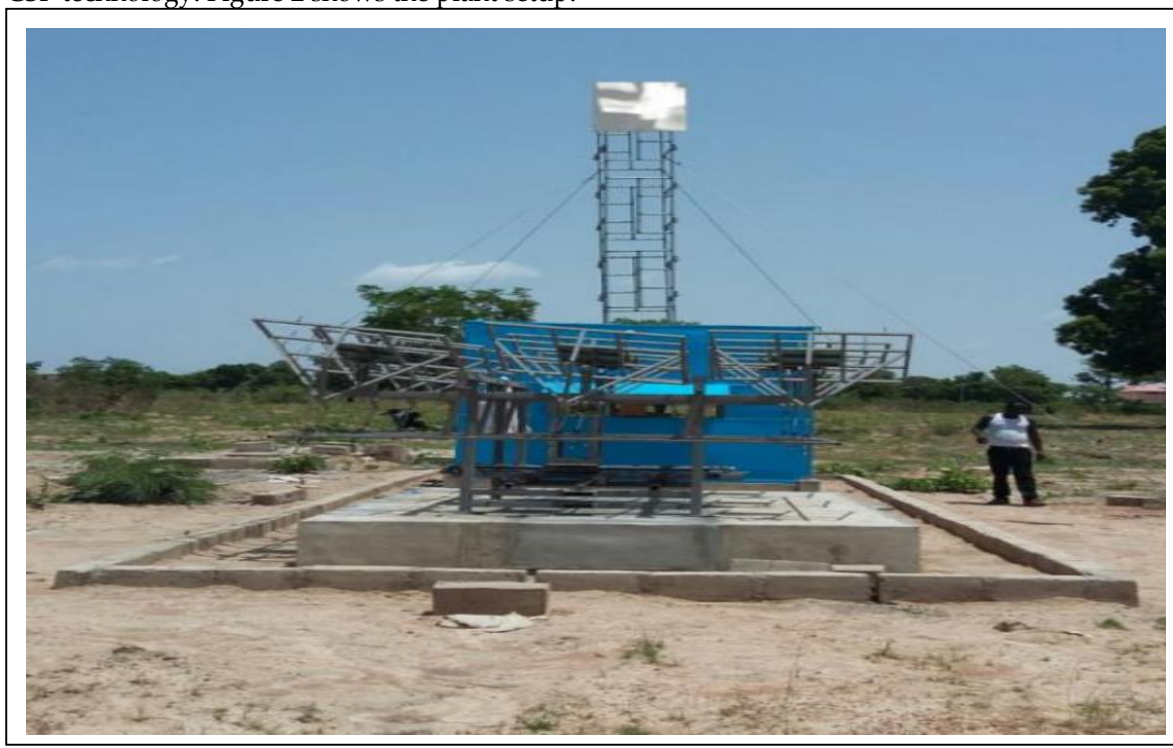

**Figure 2.** Pilot CRS micro-plant in the northern part of Ghana.

However, an assessment of the pilot TBHEC micro-CSP plant revealed that due to some design and implementation challenges of the tracking and focusing system, the plant could not operate at optimum efficiency and low cost as initially intended. The assessments have revealed that the pilot CSP's electricity generation efficiency was low because the solar tracking systems could not orient the solar panels in the right direction to capture most of the solar energy from the Sun and focus it on the receiver. Such challenges of misalignment are due to difficulties with the mechanical and control algorithms. An upgrade to project design and implementation in phase 2 solved the mechanical problem but that did little to solve the problem of misalignment, and this contributed to the realization that the key limiting factor for reaching the pilot plant's optimum operational performance was the tracking algorithm's deficiency in solving the problem of misalignment. The tracking algorithm for the pilot plant was briefly described in the design implementation report published by the authors in [7]. The solar tracking system uses a two-axis motorized system in an open-loop system. The position of the Sun was calculated with an algorithm based on the Astronomical Almanac's algorithm, which has been proposed by [32]. The tracking algorithm was then loaded on an electronic card so that a connection with a computer would not be required for its operation. The precision of the algorithm was estimated as 0.01′, which was higher than what could be expected as mechanical precision during the construction of the heliostat, but this type of tracking algorithm has a fundamental flaw of not verifying whether the reflected solar beam is accurately pointing at the collector, and initial small error margins accumulate into a big error margin unless the tracker is reset periodically. This flaw was the cause of the misalignment problem experienced in the pilot CSP project. There is, therefore, the need to devise efficient algorithms that take into consideration this flaw in subsequent phases of the project implementation.

Although the pilot micro-CSP has shown that CSP plants have the potential to reduce costs, reduce greenhouse gas emissions, and reduce the energy gap in underserved remote areas and off-grid communities in the sub-region, there is a need for further research work to broaden the acceptance and adoption of this technology in Africa. While much of the existing research work focuses on investigating the technology's techno-economic feasibility, design [33], and implementation [22] of micro-CSP plants, limited attention is given to designing and implementing precise tracking systems. However, the tracking system largely contributes to 20% of the CSP plants' operational and maintenance cost [22]. It has been established that CSP plants that have optimal tracking systems are more efficient [34]. For this reason, the remaining sections of this paper will review published literature on CSP tracking systems to highlight their strengths and weaknesses. The review aims at presenting optimal solar tracking design implementations that could be adopted to improve the efficiency of the pilot TBHEC pilot CSP plant under discussion in subsequent phases of the project implementation.

## 4. Review of Tracking Systems for CRS

Highly accurate solar tracking control and focusing systems enable CRSs to properly orient the heliostats toward the Sun and focus the majority of the incoming Sun rays onto the receiver to ensure high concentration ratios. There are two main types of CSP tracking systems depending on their freedoms/degrees of movement. First are the single-axis solar tracking systems [34,35], which track the Sun from one side to another using a single pivot point to rotate. This type of tracking system is suitable for line focusing systems. The second one that is applicable in CRS is a dual-axis solar tracking system that tracks the Sun in two different axes using two pivot points to rotate [36]. CSP tracking systems can further be grouped into active and passive systems. The former requires an external power source while the latter does not. Tracking mechanisms can also be classified into open-loop control and closed-loop control. In an open-loop control passive tracking system, the tracking system uses mathematical formulae to predict the Sun's movement without requiring any feedback [37]. The controller computes the equation into the tracking system using only the current state and the algorithm of the system and without using feedback to determine whether the output is the desired result. Commercial CRS systems use astronomical equations to determine the Sun vector for each heliostat and compute the normal vector in such a way that it divides the angle formed by the Sun vector and the vector joining the center of the heliostat to the receiver [8].

Each of these control mechanisms has advantages and disadvantages. The open-loop passive tracking system is location specific and has many sources of error, such as time, Sun model, site latitude and longitude, field heliostat distance, time-varying astigmatism, cosine effects, processor precision, atmospheric refraction, control period and structural, and mechanical and installation tolerances [38]. Moreover, the lack of feedback increases the complexity of the control system and makes the system less accurate. On the other hand, a passive closed-loop system [37] requires feedback from the Sun's position and often employs a pair of opposing solar-powered actuators that are positioned to receive equal solar radiation only when the mirrors point directly at the Sun. Misalignment with the solar vector causes negative feedback that serves to continuously track the Sun. Furthermore, some CRS tracking systems use a closed-loop active system which relies on sunlight falling on electro-optical sensors to provide alignment error feedback signal to the control system [39,40] . The feedback error is used to properly adjust the heliostats. However, the operation of the actuators depends largely on environmental factors. The various forms of CSP tracking systems are summarized in Figure 3.

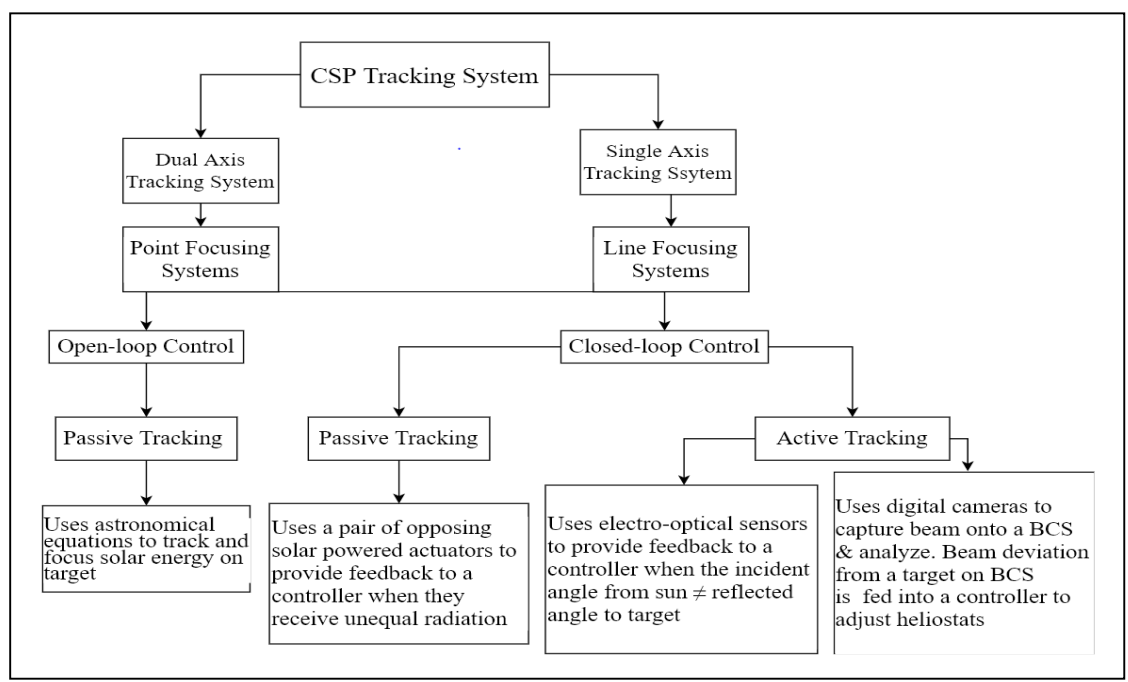

**Figure 3.** Different forms of CSP tracking systems. BCS—beam characterization system.

The performance and operational cost of every CRS plant depend largely on the accuracy of the tracking system to track the sunbeams and target them on the receiver to extract optimum energy from the Sun. Hafez et al [41]presented a comprehensive analysis of solar tracking systems, their design strategies, and the benefits in solar application. The survey revealed innovative and cost-effective methods for developing CRS tracking systems to optimize the quantity and quality of solar energy captured from the Sun. The tracking system's operating cost and the complexity associated with the design of the tracking system were highlighted as the main challenges of CRS tracking systems. For a similar objective, Vermaak et al [42] analyzed the operational costs of tracking systems and their outputs in terms of energy production under four different configurations and three different scenarios. The authors established that the dual-axis tracking system produced 3% higher energy than the single-axis tracking system and 22.28% higher when compared to a fixed PV system. The authors concluded that tracking systems are not always cost effective and therefore the use of each tracking system should be assessed carefully before. These review papers afforded us information about the various tracking systems in solar energy production.

In commercial CSP systems, open-loop passive tracking systems are mainly used due to the simplicity in their design and implementation. However, their application in CRS plants often leads to less harvesting of solar energy due to many errors. To resolve some of these errors, Yeguang Hu et al [43]proposed a new tracking system that reduces cosine losses and improves concentration efficiency through the use of the principles of coordinate rotation transformation and non-imaging optics. The proposed algorithm was simulated, and the results showed an improved concentration ratio as compared to the open-loop tracking systems. However, this system requires the heliostats to be very close to the receiver. Electro-optical sensors can be used to make tracking systems more efficient. Electro-optical sensor-based tracking systems measure the incident and reflected angles on the heliostat relative to the normal to ensure that these angles are always the same for the law of reflection to be obeyed. Any deviation is reported as an error and used in heliostat adjustment. Camacho and Berenguel [44] provided studies on the heliostat reflection imaging mechanisms and the relationship among these angles and the control period. The authors demonstrated that advanced control techniques can be used to increase CRS plant performance.

Some commercial CRS solar tracking systems combine open-loop tracking systems with Sun-sensor methods [45]. or digital cameras [46] The digital cameras provide every heliostat with local error feedback by imaging the reflected angle of the beam. Pfahl et al in patent paper [47] proposed a method to measure relative angles between the heliostat normal, the receiver, and the Sun using the heliostat-mounted imaging device. Kribus et al [48] described a scheme that directly measures individual heliostat errors by "looking back" from the receiver to the field using cameras. However, sensors do not provide accurate measurements due to environmental factors, and the use of cameras for each central processor leads to extensive image processing if the algorithm is not optimal. As a result, Zhu Xuemei et al [49] has developed a nonlinear kinematic model that uses a beam characterization system (BCS) and a digital camera. The camera captures the reflected solar image on the calibration target (BCS) located on the tower below the receiver, and the motion trajectory information of a single heliostat is sampled on the target to form the Sun's image reference point. A deviation from this reference point, whenever a sample is taken, serves as a feedback that is used to adjust the heliostat to track the Sun and properly focus the reflected rays on the receiver. Figure 4 shows the schematic diagram of the proposed system.

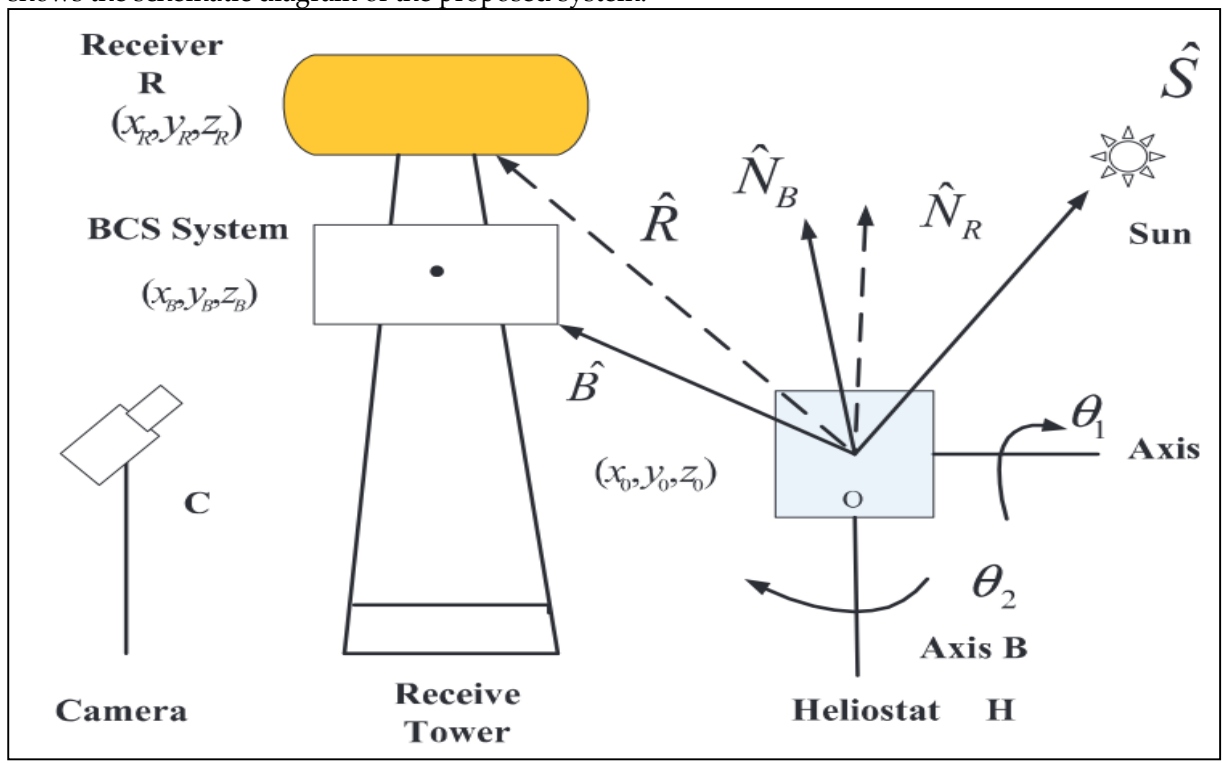

**Figure 4.** Heliostat calibration system with BCS and machine version [48].

Using the same concept, Berenguel et al [50] proposed a model to correct the calculation and installation of the solar position. The model is based on the use of a charge-device camera to capture images of the reflected beams from each field heliostat to the target to be used for offset correction. Although the models produced good simulation results, they are computationally expensive and costly to implement in the practical application of the CRS.

Salgado-plasencia et al [51] proposed an improved heliostat control system based on the SCADA system. The system uses fuzzy logic to calculate the position of the Sun and the desired angles using the Solar Position Algorithm and Control Unit. The SCADA monitors and controls the orientation of the solar angles. However, the authors tested only the proposed system to assess the performance of the remote terminal unit. With the advent of optimal machine vision algorithms, recent CRS tracking systems combine the open-loop control system with the machine vision for error correction to increase the efficiency of the control system or machine vision on its own to implement the tracking system [36,45,52]. For example, Burisch et al [53] proposed an automatic calibration procedure for each heliostat attached to a digital camera. The camera is used to observe different targets on the receiver under different orientations of the heliostat. When optimum heliostat orientation is observed, the parameters describing the motion of the heliostat are estimated and calibrated. Guangyu and Zhongkun [54] also proposed a closed-loop feedback control method based on the machine vision and optical reflection principle. The proposed system captures reflections from the target and uses image processing to detect and extract solar angles based on BP neural networks. Carballo et al [55] developed a low-cost machine vision solar tracking system that can be used for any type of solar system. The proposed system calculates the Sun vector, the target vector of the receiver and the aiming point vector on a camera plane below the receiver tower. All these vectors are computed relative to the origin of the individual heliostats, and a deviation from the target when a sample is taken constitutes an error that is fed as an input into the control to adjust the heliostat to appropriate position. Figure 5 shows the schematic diagram of the proposed tracking system. The tracking system provides monitoring and control of the orientation of the solar angles as well as cloud movement prediction, block and shadow detection, atmospheric attenuation, or measures of concentrated solar radiation, which can be used to improve the control strategies of the system and its performance. The proposed system was tested on Plataforma solar de Almer´ıa (PSA) [15], and the results showed great performance improvement over traditional tracking systems. However, its applicability has not been reported. The strengths and weaknesses of the tracking systems reviewed are discussed in the next section.

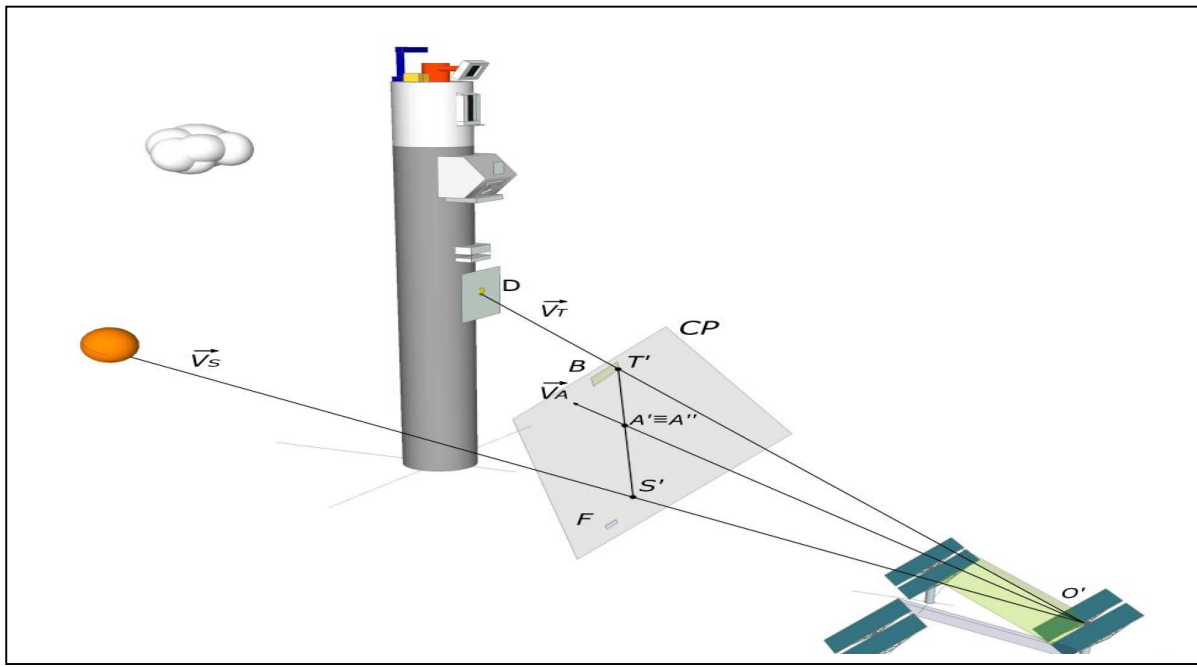

**Figure 5.** Solar tracking system based on computer vision [54].

## 5. Discussion

Solar tracking systems are responsible for guaranteeing the correct daily operation of the solar plant by computing the tracking setpoint and managing communications among the various heliostats, the receiver, and the control drive. Solar tracking systems are also responsible for performing error diagnosis and correction and making decisions in emergencies. Key among these duties of a solar tracking system is its ability to minimize the following while consuming a minimal amount of the harvested energy:

- Aiming errors: These errors are related to constant aberrations including pedestal tilt due to displacement of the center of gravity, bad reference, or structure deformations due to wind loads and solar radiation refraction.

- Tracking offset errors: These errors are related to the tracking system's movement, the resolution of the movement, and the controller's accuracy.

- Optical errors that can be attributed to astigmatic aberration.

Table 1 grades the tracking systems under two different performance metrics. Metric one considers the computational and energy cost of the control algorithm, and metric two considers the ways by which the tracking system is prone to errors and may not function efficiently.

**Table 1.** Performance metrics of the different types of CRS tracking systems.

| Types of CRS Tracking systems | Reference | Cost of Control system | Errors in tracking System | Comment |
|---|---|---|---|---|
| Open-loop Tracking Systems | Michalsky [32] | Medium | High | Cannot verify whether reflected solar beam is accurately pointing at collector and initial small error margin accumulates into a big error margin unless the tracker is reset periodically. |
| Sensor-Based Tracking Systems | Andreas Pfahl et al. [47], Lastoviet al. [45] | Medium | Medium | Cannot eliminate astigmatism and cosine errors and errors associated with computing the three vectors to obtain the incident and reflected angles. Accuracy also depends on environmental factors, controller, and sensor resolution. |
| Machine vision-Based Tracking Systems with Camera behind the target | Kribus et al. [48] | High | High for a large number of heliostats | Computationally expensive and when several reflected images are captured, it becomes difficult to distinguish among which of the heliostats are directly focusing solar energy on the receiver and which ones should be adjusted. |
| Machine vision Based Tracking Systems with BCS below target | Zhu Xuemei et al. [49], Berenguel et al. [50], Mohamed Adel et al. [46], Guangyu and Zhongkun [54], Carballo et al. [55] | High | Medium | Computationally expensiveness, errors from beam characterization and heliostat control as well as communication delays make this tracking system less suitable. |
| Other mechanisms of CRS tracking systems | Eugenio et al. [51], Yeguang Hu et al. [43] | High | High | Requires heliostat to be closed to the target, requires a two-stage control system that needs to highly synchronize, and this decreases the accuracy of the system. The system is not suitable for a large CRS plant. |

One major challenge for the pilot plant CRS system deployed in the northern part of Ghana was its inability to simultaneously increase the accuracy of the control system and reduce its computational costs. Open-loop passive tracking systems have low error margins and are less computationally expensive, but from the literature summarized in Table 1, these types of tracking systems have a high error rate. Errors are the result of constant aberrations and lack of feedback. Lack of feedback means that the errors accumulate over some time cannot be detected and corrected automatically. Due to this limitation, some commercial CRS tracking systems implement this type of electro-optical sensor tracking system so that the sensors provide feedback to the control system. Electro-optical sensor-based systems are independent of time and location. They also have medium costs and errors. However, they can only obtain a relative solar position and control the solar system when the optical axis of the system is aligned with the Sun vector. Thus, the accuracy of these systems is mostly influenced by the resolution of the controller and sensor, and their initial installation is also error prone, making their operation less efficient. Moreover, electro-optical sensors are easily affected by adverse environmental factors that cause false readings. Furthermore, there are time-varying astigmatism and cosine errors and many errors associated with the geometric incident and reflecting angles at the center of the heliostat. All of these errors limit the accuracy of sensor-based CRS tracking systems.

A much better approach to designing more nearly accurate CRS tracking systems employs machine vision where digital cameras replace the optical sensors. The cameras are used to capture the sunbeam onto a beam characterization system (BCS) placed below the target. The beam is then analyzed to obtain the deviation from the target point on the BCS. The error is then fed into the control system to adjust the heliostat appropriately. This approach shows improved operational efficiency but has a high cost of running the control algorithm. It also requires complex signal processing or training algorithms to produce an accurate control system, and when this approach is coupled with the centralized control system to control thousands of heliostats, it further increases the complexity and energy consumption of the tracking system. Another major challenge of this approach is the control mechanism. Some tracking systems centralize the control system while others cluster it. Either way requires the control system to continuously perform complex signal processing to characterize the beams and adjust the heliostats. Additionally, the communication delay between the control systems and the heliostats could contribute greatly to tracking errors as indicated in Table 1.

Based on the study, it can be concluded that future tracking systems design and upgrades to the pilot plant should focus on increasing the efficiency of the CSP plant and reducing the operational cost of the tracking system by using algorithms that use neural artificial intelligence machine vision to characterize the beam and optimally re-adjust the pilot plant's multifaceted heliostats. This research is currently in progress and novel algorithms using machine learning and artificial intelligence are being developed and tested to solve the misalignment of the pilot micro-CSP plant's multifaceted heliostats.

## 6. Conclusion and Recommendation

The paper highlighted the energy situation and current renewable energy research development in Africa. Technologies required to bridge the energy deficit of sub-Saharan African countries are discussed. It has been noted that some African countries are working on the possibility of using CSP technology to reduce their respective countries' energy deficit. Among other cost–performance advantages of CSP technology, tracking systems contribute significantly to higher operating temperatures and higher-efficiency power generation. This contribution comes from innovation in the control algorithm design of the tracking system which aims to reduce its running costs and maximize its performance. The major tracking systems for a pilot micro-CSP plant have been presented in this review paper. The review revealed that CSP tracking systems can be broadly categorized as either dual-axis or single-axis tracking systems and closed-loop or open-loop tracking systems based on their degree of movement and control mode, respectively. The principles of control and computation of each system were discussed, and the efficiency and relative advantages and drawbacks were provided. The findings reported in this review validate the explanations why the pilot micro-CSP plant implemented by TBHEC for an off-grid community in Ghana does not run at full efficiency. Since there is an increasing interest in applying machine learning and artificial intelligence to design CSP tracking systems to improve their performance, it is our view that novel algorithms with less computational complexity be tested on the pilot plant to evaluate its performance. The challenge, however, is the complex image processing requirements of machine learning. We, therefore, recommend that future research works should focus on reducing these requirements for the control algorithm of CSP tracking systems to increase their operational efficiency.

**Author Contributions:** conceptualization, E. W. R. and E. T. T.; methodology, E. W. and E. T. T.; validation, J. J. K.; formal analysis, E. T. T.; investigation, Y. A. K. F.; resources, Y. A. K. F.; data curation, A. S. A.; writing—original draft preparation, A. S. A.; writing—review and editing, Y. A. K. F.; supervision, J. J. K.; project administration, E. W. R.; funding acquisition, E. W. R.

**Funding:** The APC was funded by Office of Grants Research (OGR) under the auspices of the Kwame Nkrumah University of Science and Technology Research Fund (KReF).

**Conflicts of Interest:** The authors declare no conflict of interest.

**References**

1. Nitz, P.; Fluri, T.; Lude, S.; Meyer, R.; Alasis, E.; Tawalbeh, M.; Shahin, W. On the Way to the First CSP Pilot Plant in Jordan: The WECSP Project. *Energy Procedia* **2015**, *69*, 1652–1659.
2. Kassai, M. Experimental investigation of carbon dioxide cross-contamination in sorption energy recovery wheel in ventilation system. *Build. Serv. Eng. Res. Technol.* **2018**, *39*, 463–474.
3. Martins, F.; Felgueiras, C.; Smitkova, M.; Caetano, N. Analysis of fossil fuel energy consumption and environmental impacts in european countries. *Energies* **2019**, *12*, 1–11.
4. World Bank. *The Word Bank—Annual Report 2017*; World Bank: Washington, DC, USA, 2017.
5. Jan Corfee-Morlot, F.A.; Parks, P.; Ogunleye, J. *Achieving Clean Energy access in Sub-Saharan Africa a Case Study for the OECD, UN Environment, World Bank Project: Financing Climate Futures: Rethinking Infrastructure*; World Bank: Washington, DC, USA, 2018.
6. Ogunmodimu, O.; Okoroigwe, E.C. Concentrating solar power technologies for solar thermal grid electricity in Nigeria: A review. *Renew. Sustain. Energy Rev.* **2018**, *90*, 104–119.
7. Seshie, Y.M.; Tsoukpoe, K.E.N.; Neveu, P.; Coulibaly, Y. Small scale concentrating solar plants for rural electri fi cation. *Renew. Sustain. Energy Rev.* **2018**, *90*, 195–209.
8. Irena, W.D.I. *UNSD, WB, WHO Tracking SDG 7: The Energy Progress Report 2019*; UNSD, WB, WHO: Geneva, Switzerland, 2019.
9. Xu, M.; Zhu, X. Simulation and Control of Heliostat Sun-tracking in Central Receiver Solar Power Plant. In Proceedings of the 2018 37th Chinese Control Conference, Wuhan, China, 25–27 July 2018; Volume 2, pp. 7553–7557.
10. Pfahl, A.; Coventry, J.; Röger, M.; Wolfertstetter, F.; Vásquez-Arango, J.F.; Gross, F.; Arjomandi, M.; Schwarzbzl, P.; Geiger, M.; Liedke, P. Progress in heliostat development. *Sol. Energy* **2017**, *152*, 3–37.
11. Abdollahpour, M.; Golzarian, M.R.; Rohani, A.; Zarchi, H.A. Development of a machine vision dual-axis solar tracking system. *Sol. Energy* **2018**, *169*, 136–143.
12. Islam, R.; Bhuiyan, A.B.M.N.; Ullah, M.W. An Overview of Concentrated Solar Power (CSP)Technologies and its Opportunities in Bangladesh. In Proceedings of the 2017 International Conference on Electrical, Computer and Communication Engineering, Cox's Bazar, Bangladesh, 16–18 February 2017; pp. 844–849.
13. Zhu, G.; Libby, C. Review and future perspective of central receiver design and performance. In *AIP Conference Proceedings*; AIP Publishing LLC: Melville, NY, USA, 2017; Volume 1850.
14. Viebahn, P.; Lechon, Y.; Trieb, F. The potential role of concentrated solar power (CSP) in Africa and Europe-A dynamic assessment of technology development, cost development and life cycle inventories until 2050. *Energy Policy* **2011**, *39*, 4420–4430.
15. Masood, R.; Gilani, S.I.U.H.; Al-Kayiem, H.H. A simplified design procedure of parabolic trough solar field for industrial heating applications. *ARPN J. Eng. Appl. Sci.*, **2016**, *11*, 13065–13071.
16. Ravelli, S.; Franchini, G.; Perdichizzi, A.; Rinaldi, S.; Valcarenghi, V.E. Modeling of Direct Steam Generation in Concentrating Solar Power Plants. *Energy Procedia* **2016**, *101*, 464–471.
17. Meligy, R.; Rady, M.; el Samahy, A.; Mohamed, W.; Paredes, F.; Montagnino, F. Simulation and control of linear Fresnel reflector solar plant. *Int. J. Renew. Energy Res.* **2019**, *9*, 804–818.
18. Malan, K.; Gauche, P. Model based open-loop correction of heliostat tracking errors. *Energy Procedia* **2013**, *49*, 7.
19. Stein, K.L.W. *Concentrating Solar Power Technology: Principles, Developments and Applications*; Woodhead Publishing Limited: Cambridge, UK, 2012.
20. Sonawane, P.D.; Raja, V.K.B. An overview of concentrated solar energy and its applications. *Int. J. Ambient. Energy* **2018**, *39*, 898–903.
21. Leiva-Illanes, R.; Escobar, R.; Cardemil, J.M.; Alarcón-Padilla, D.C.; Uche, J.; Martínez, A. Exergy cost assessment of CSP driven multi-generation schemes: Integrating seawater desalination, refrigeration, and process heat plants. *Energy Convers. Manag.* **2019**, *179*, 249–269.
22. N'Tsoukpoe, K.E.; Azoumah, K.Y.; Ramde, E.; Fiagbe, A.K.Y.; Neveu, P.; Py, X.; Gaye, M.; Jourdan, A. Integrated design and construction of a micro-central tower power plant. *Energy Sustain. Dev.* **2016**, *31*, 1–13.
23. Balghouthi, M.; Trabelsi, S.E.; Amara, M.B.; Ali, A.B.H.; Guizani, A. Potential of concentrating solar power (CSP) technology in Tunisia and the possibility of interconnection with Europe. *Renew. Sustain. Energy Rev.*

**2016**, *56*, 1227–1248.
24. Zlatanov, H.; Weinrebe, G. CSP and PV solar tracker optimization tool. *Energy Procedia* **2014**, *49*, 1603–1611.
25. Aly, A.; Bernardos, A.; Fernandez-Peruchena, C.M.; Jensen, S.S.; Pedersen, A.B. Is Concentrated Solar Power (CSP) a feasible option for Sub-Saharan Africa? Investigating the techno-economic feasibility of CSP in Tanzania. *Renew. Energy* **2019**, *135*, 1224–1240.
26. Belgasim, B.; Aldali, Y.; Abdunnabi, M.J.R.; Hashem, G.; Hossin, K. The potential of concentrating solar power (CSP) for electricity generation in Libya. *Renew. Sustain. Energy Rev.* **2018**, *90*, 1–15.
27. Pan, C.A.; Dinter, F. Combination of PV and central receiver CSP plants for base load power generation in South Africa. *Sol. Energy* **2017**, *146*, 379–388.
28. Mihoub, S.; Chermiti, A.; Beltagy, H. Methodology of determining the optimum performances of future concentrating solar thermal power plants in Algeria. *Energy* **2017**, *122*, 801–810.
29. Bouhal, T.; Agrouaz, Y.; Kousksou, T.; Allouhi, A.; El Rhafiki, T.; Jamil, A.; Bakkas, M. Technical feasibility of a sustainable Concentrated Solar Power in Morocco through an energy analysis. *Renew. Sustain. Energy Rev.* **2018**, *81*, 1087–1095.
30. Elshafey, S. Solar thermal power in Egypt. In Proceedings of the 2018 IEEE Industrial Applications Society Annual meeting, IAS 2018, Portland, OR, USA, 23–27 September 2018; pp. 1–8.
31. Craig, T.; Brent, A.; Duvenhage, F.; Dinter, F. Systems approach to concentrated solar power (CSP) technology adoption in South Africa. In *AIP Conference Proceedings*; AIP Publishing LLC: Melville, NY, USA, 2018; Volume 2033.
32. Michalsky J. J. The astronomical almanac's algorithm for approximate solar position (1950–2050). Sol Energy 1988, *40*, 227–35.
33. Zoschke, T.; Frantz, C.; Schöttl, P.; Fluri, T.; Uhlig, R. Techno-economic assessment of new material developments in central receiver solar power plants. In *AIP Conference Proceedings*; AIP Publishing LLC: Melville, NY, USA, 2019; Volume 2126.
34. Jafrancesco, D.; Cardoso, J.P.; Mutuberria, A.; Leonardi, E.; Les, I.; Sansoni, P.; Francini, F.; Fontani, D. Optical simulation of a central receiver system: Comparison of different software tools. *Renew. Sustain. Energy Rev.* **2018**, *94*, 792–803.
35. Ghassoul, M. Single axis automatic tracking system based on PILOT scheme to control the solar panel to optimize solar energy extraction. *Energy Rep.* **2018**, *4*, 520–527.
36. Kamal, M. Microcontroller Based Single Axis Sun Tracking Control System. *Int. J. Emerg. Technol. Eng. Res.* **2018**, *3*, 3–7, .
37. Garcia-Gil, G.; Ramirez, J.M. Fish-eye camera and image processing for commanding a solar tracker. *Energies* **2019**, *5*, e01398.
38. Lee, C.Y.; Chou, P.C.; Chiang, C.M.; Lin, C.F. Sun tracking systems: A review. *Sensors* **2009**, *9*, 3875–3890.
39. Stone, C.W.; Lopez, K. *Evaluation of the Solar One Track Alignment Methodology*; American Society of Mechanical Engineers: New York, NY, USA, 1995; p. 1995.
40. Sansoni, P.; Fontani, D.; Francini, F.; Jafrancesco, D.; Mercatelli, L.; Sani, E. Pointing sensors and sun tracking techniques. *Int. J. Photoenergy* **2011**, *2011*, 25.
41. Hafez, A.Z.; Yousef, A.M.; Harag, N.M. Solar tracking systems: Technologies and trackers drive types—A review. *Renew. Sustain. Energy Rev.* **2018**, *91*, 754–782.
42. Vermaak, H.J. Techno-economic analysis of solar tracking systems in South Africa. *Energy Procedia* **2014**, *61*, 2435–2438.
43. Hu, Y.; Shen, H.; Yao, Y. A novel sun-tracking and target-aiming method to improve theconcentration efficiency of solar central receiver systems. *Renew. Energy* **2017**, *120*, 17.
44. Camacho, E.F.; Berenguel, M. Control of solar energy systems. *IFAC Proc.* **2012**, *8*, 848–855.
45. Laštovi, G. Open Source Sun Tracking System with Solar Panel Monitoring and Heliostat Control System. *Int. J. Contemp. Energy* **2017**, *3*, 44–50.
46. Adel, M.; Peña-Lapuente, A.; Rady, M.; Hamdy, A. Development of a New High Dynamic Range Technique for Solar Flux Analysis Using Double-CCD Optical Camera. *Int. J. Renew. Energy Res.* **2019**, *9*, 783–794.
47. Pfahl, R. A.; Karsten, R.B. Method for Controlling the Alignment of A Helostat with Respect To A Receiver, Heliostat Device and Solar Power Plant. U.S. Patent 8,651,100, 18 February 2014.
48. Kribus, A.; Vishnevetsky, I.; Yogev, A.; Rubinov, T. Closed loop control of heliostats. *Energy* **2004**, *29*, 905–913.

49. Xuemei, Z.; Xiaoling, M.; Kaizhi, L.; Wenjun, H. Precise Sun-tracking Control of Heliostats Based on a Sun's Image Reference System. In Proceedings of the 26th Chinese Control and Decision Conference, Changsha, China, 31 May–2 June 2014; Volume 14, pp. 3–6.
50. Berenguel, M.; Rubio, F.R.; Valverde, A.; Lara, P.J.; Arahal, M.R.; Camacho, E.F.; López, M. An artificial vision-based control system for automatic heliostat positioning offset correction in a central receiver solar power plant. *Sol. Energy* **2004**, *76*, 563–575.
51. Salgado-plasencia Eugenio .; Carrillo-serrano, R.V.; Rivas-araiza, E.A.; Toledano-ayala, M. SCADA-Based Heliostat Control System with a Fuzzy Logic Controller for the Heliostat Orientation. *Appl. Sci.* **2019**, *9*, 20.
52. Sohag, H.A.; Hasan, M.; Khatun, M.M.; Ahmad, M. An Accurate and Efficient Solar Tracking System Using Image Processing and LDR Sensor. In Proceedings of the 2015 2nd International Conference on Electrical Information and Communication Technologies, Khulna, Bangladesh, 10–12 December 2015; pp. 522–527.
53. Burisch, M.; Sanchez, M.; Olarra, A.; Villasante, C. Heliostat calibration using attached cameras and artificial targets. In *AIP Conference Proceedings*; AIP Publishing LLC: Melville, NY, USA, 2016; Volume 1743.
54. Guangyu, L.I.U.; Zhongkun, C.A.I. Heliostat attitude angle detection method based on BP neural network. In *MATEC Web of Conferences*; EDP Sciences: Julius, France, 2017; Volume 139.
55. Carballo, J.A.; Bonilla, J.; Berenguel, M. New approach for solar tracking systems based on computer vision, low cost hardware and deep learning. *Renew. Energy* **2018**, *133*, 1158–1166.